\documentclass[twocolumn,prb,superscriptaddress,floatfix]{revtex4-2}
\usepackage[T1]{fontenc}
\usepackage[utf8]{inputenc}
\usepackage{color}
\usepackage{units}
\usepackage{amssymb}
\usepackage{graphicx}
\usepackage{esint}
\usepackage{bm}
\usepackage{xcolor, braket}
\usepackage[breaklinks=true,colorlinks=true,urlcolor=blue,
citecolor=blue,linkcolor=black,bookmarks=false]{hyperref}
\usepackage{setspace}
\newcommand{\rxx}{\ensuremath{R_\mathrm{xx}} }
\newcommand{\rxy}{\ensuremath{R_\mathrm{xy}} }

\newcommand{\bperp}{\ensuremath{B_\perp} }
\newcommand{\bpar}{\ensuremath{B_\parallel} }
\newcommand{\bx}{\ensuremath{B_x} }
\newcommand{\by}{\ensuremath{B_y} }
\newcommand{\rb}{\ensuremath{\rxx(\bpar)} }

\begin{document}

\title{Metastable magnetic domains and the anomalous $B_\parallel=0$ resistance peak in twisted double bilayer graphene}

\author{Zhenxiang Gao}
\thanks{These authors contributed equally}
\affiliation{Quantum Matter Institute, University of British Columbia, Vancouver, British Columbia, V6T 1Z1, Canada}
\affiliation{Department of Physics and Astronomy, University of British Columbia, Vancouver, British Columbia, V6T 1Z1, Canada}

\author{Christopher Coleman}
\thanks{These authors contributed equally}
\affiliation{Quantum Matter Institute, University of British Columbia, Vancouver, British Columbia, V6T 1Z1, Canada}
\affiliation{Department of Physics and Astronomy, University of British Columbia, Vancouver, British Columbia, V6T 1Z1, Canada}

\author{Silvia L\"uscher}
\affiliation{Quantum Matter Institute, University of British Columbia, Vancouver, British Columbia, V6T 1Z1, Canada}
\affiliation{Department of Physics and Astronomy, University of British Columbia, Vancouver, British Columbia, V6T 1Z1, Canada}

\author{Ruiheng Su}
\affiliation{Quantum Matter Institute, University of British Columbia, Vancouver, British Columbia, V6T 1Z1, Canada}
\affiliation{Department of Physics and Astronomy, University of British Columbia, Vancouver, British Columbia, V6T 1Z1, Canada}

\author{Manabendra Kuiri}
\affiliation{Quantum Matter Institute, University of British Columbia, Vancouver, British Columbia, V6T 1Z1, Canada}
\affiliation{Department of Physics and Astronomy, University of British Columbia, Vancouver, British Columbia, V6T 1Z1, Canada}
\affiliation{Department of Physics, Birla Institute of Technology and Science, Pilani, Hyderabad Campus, Telangana 500078, India}

\author{Kenji Watanabe}
\affiliation{Research Center for Electronic and Optical Materials, National Institute for Materials Science, 1-1 Namiki, Tsukuba 305-0044, Japan}

\author{Takashi Taniguchi}
\affiliation{Research Center for Materials Nanoarchitectonics, National Institute for Materials Science, 1-1 Namiki, Tsukuba 305-0044, Japan}

\author{Nemin Wei}
\affiliation{Department of Physics, Yale University, New Haven, Connecticut, USA}

\author{Chunli Huang}
\affiliation{Department of Physics and Astronomy, University of Kentucky, Lexington, Kentucky 40506-0055, USA}

\author{Joshua Folk}
\email{jfolk@physics.ubc.ca}
\affiliation{Quantum Matter Institute, University of British Columbia, Vancouver, British Columbia, V6T 1Z1, Canada}
\affiliation{Department of Physics and Astronomy, University of British Columbia, Vancouver, British Columbia, V6T 1Z1, Canada}

\date{\today}

\begin{abstract}
In graphene moir\'es, valley polarization gives rise to orbital magnetism, manifested as an anomalous Hall effect and resulting in Barkhausen jumps in longitudinal resistance when changing domain configurations modify quasiparticle scattering. Beyond a simple picture of polarized domains, however, spin and valley textures within and between the domains are less well understood, as is the effect of these textures on transport. In the valley-polarized quarter-metal state of twisted double bilayer graphene, a sharp and metastable peak in longitudinal resistance often appears at zero in-plane magnetic field, whose microscopic origin has yet to be identified.  Here, we show that this peak depends on the configuration of domains of orbital magnetism, which is itself set by the gate-voltage trajectory used to enter the ordered state and by the magnetic field --- particularly the in-plane component --- present during that trajectory.  The sensitivity of the effect to in-plane magnetic field components points to spin, linked to valley polarization through spin-orbit coupling, as the key degree of freedom in both the domain formation and the resistance peak.
\end{abstract}

\maketitle

\section{Introduction}

Twisted double bilayer graphene (t2+2) --- two Bernal-stacked bilayers with a relative twist --- belongs to a family of twisted multilayer graphene systems in which the moir\'e bands are tunable not only by twist angle but also by out-of-plane displacement field.\cite{waters2024topological}  For twist angles between roughly $1^\circ$ and $1.4^\circ$, the t2+2 conduction band can be made flat enough for electron-electron interactions to spontaneously break its intrinsic four-fold spin-valley degeneracy.  Under these conditions, correlated insulating states appear at integer fillings $\nu$ of the moir\'e unit cell, and between integer fillings the system polarizes into a subset of the four spin-valley flavours.\cite{lee2019theory,shen2020correlated,burg2019correlated,liu2020tunable,cao2020tunable,he2021symmetry,kuiri2022spontaneous,liu2022isospin,he2023symmetry,liu2023observation}

In t2+2, the correlated insulator at $\nu=2$ is surrounded by a broad region of elevated resistance --- the `halo' --- in which Hall measurements indicate that two of the four spin-valley degeneracies are spontaneously broken.\cite{liu2020tunable,cao2020tunable,he2021symmetry,liu2022isospin,kuiri2022spontaneous,liu2023observation}  This two-fold-degenerate state, often called a half-metal, is has been frequently reported to be spin-polarized but valley-degenerate.  Near the tip of the halo, at fillings above $\nu\approx 3.5$, the remaining valley degeneracy is lifted and the system becomes a quarter-metal.  Because the two valleys carry opposite Berry curvature, valley polarization in the quarter-metal produces a net orbital magnetization and an anomalous Hall effect (AHE).  In Ref.~\cite{kuiri2022spontaneous}, several of the authors of the present paper reported an AHE above $\nu=3$ in t2+2, confirming valley polarization at those fillings.  The AHE was shown to be orbital in origin by its extreme anisotropy with respect to magnetic field direction: the hysteresis is driven entirely by $\bperp$, with $\bpar$ having no measurable effect even up to several Tesla.

A significant puzzle left unexplored in Ref.~\cite{kuiri2022spontaneous} was the fact that, although relatively large values of $\bpar$ affected the AHE only weakly, as observed in the transverse (Hall) resistance, a large but extremely narrow (few milliTesla wide) peak in longitudinal $\bpar$ magnetoresistance was consistently observed.  Notably, this peak was unstable in time: sweeping back and forth over it, the peak resistance would often jump up or down between sweeps, or even during the sweep.  From an experimental point of view, this made systematic investigation difficult, since successive measurements at nominally identical settings could yield qualitatively different results.  From a microscopic point of view, the instability hinted that metastable domain orientations might be involved.

Here, we investigate the low-field $\bpar$ magnetoresistance peak in t2+2 in detail, focusing on how this feature depends on the prior history of the electronic state in the sample. The presence or absence of the $\bpar$ peak is found to depend sensitively on the precise value of the magnetic field when gate voltages are swept into the symmetry-broken region, as well as the path taken by the sweep in gate-voltage-space.   Together, these data provide strong evidence that the $\bpar$ peak depends on the domain structure of the orbital magnetic state in t2+2.

\begin{figure*}
	\includegraphics{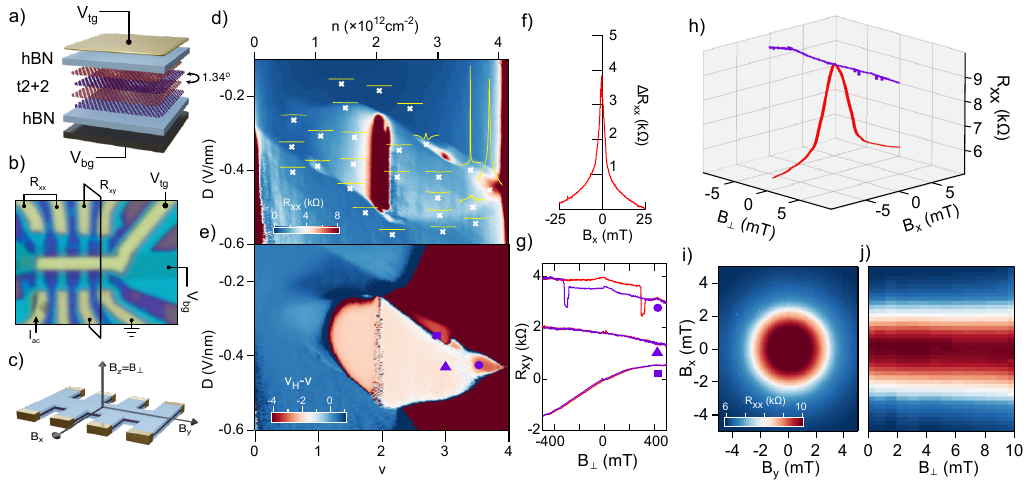}
	\caption{a) Stacking configuration of the device. t2+2 consists of two Bernal (AB)-stacked bilayer graphene with a twist angle $\theta\sim 1.34^\circ$, $V_{tg}$ is the gate voltage on a $CrAu$ top gate; $V_{bg}$ is the gate voltage applied to a graphite bottom gate. b) Optical image of the device showing measurement configuration. c) Hall bar geometry defining $B_x$ and $B_y$ as $B_\parallel$ components as well as $B_z\equiv B_\perp$. d) Longitudinal resistance $\rxx$ across the halo region (100~mK).  Yellow lines show $R_{xx}(B_x)$ measurements at the $\nu,D$ locations specified by white crosses (25~mK). e) Reduced Hall filling $|\nu_H-\nu|$ (see methods) extracted at $|\bperp|=0.5$~T across the halo region, with markers at the three $\nu,D$ locations where data in panel g) is collected (300~mK). f) Example of the $R_{xx}(B_x)$ curve at the location $\nu=3.56, D=-0.45~{\rm V/nm}$ in panel d) (100~mK). g) Transverse resistance measurements $R_{xy}(\bperp)$ at the locations marked in e) (300~mK). h)-j) The comparison between effects of $B_x$, $B_y$, and $\bperp$ on $R_{xx}$ (100~mK). h) Contrast between in-plane ($R_{xx}(B_x)$, red) and out-of-plane ($R_{xx}(\bperp)$, purple) magnetoresistance. i) $R_{xx}(\bpar)$ is isotropic within experimental resolution within the $B_x, B_y$ plane. j) In contrast, $R_{xx}(B_x)$ sweeps are unaffected by $\bperp$ up to at least 10~mT.} 
	\label{fig1}
\end{figure*}

\section{Experimental methods}

The measurements reported here were carried out on a micron-scale Hall bar composed of two Bernal stacked bilayers with a twist angle $\theta\sim 1.34^\circ$ between them, illustrated in Figs.~\ref{fig1}a-c.  This device was one of those investigated in Ref.~\cite{kuiri2022spontaneous}; although the second device in Ref.~\cite{kuiri2022spontaneous} showed an analogous $\bpar$ peak, it was damaged before history-dependent measurements of the type presented here could be completed.  Gates above and below the stack enabled independent tuning of the carrier density, $n=(C_{bg}V_{bg}+C_{tg}V_{tg})/e$, and transverse electric field, $D=(C_{bg}V_{bg}-C_{tg}V_{tg})/2\epsilon_0$, via top/bottom-gate voltages $V_{tg/bg}$.  Here $e$ is the electron charge, $\epsilon_0$ is the permittivity of free space, and $C_{tg/bg}$ are the capacitances of top/bottom gate dielectrics.  For easier interpretation,  $n$ may be scaled to the number of electrons per moire site, also referred to as the filling $\nu$. Gate voltages were independently controlled by synchronized voltage sources enabling arbitrary ramps through the 2D gate voltage plane.  Resistance measurements, $\rxx$ or $\rxy$, were made using a lock-in amplifier with a typical AC current bias of $I_{ac}=1.4$~nA.  Measurements were performed in a dilution refrigerator where the base electron temperature was measured to be around 25~mK. Hall data were collected at 300~mK, 2D gate maps of longitudinal resistance were collected at 25~mK, and the rest of the data were collected at 100~mK, except where noted.

The refrigerator was equipped with a vector magnet that allowed full 3D control of the magnetic field (Fig.~\ref{fig1}c).  Trapped flux is known to create mT-scale offsets in the magnetic field of cryogen-free magnets, depending on previous sweep history.  In this experiment, we calibrate the offset where possible using resistance features sharply centered at zero field: the $\bpar$ resistance peak that is the subject of this paper, and the $\bperp$ resistance peak due to weak localization that can be observed in gate voltage settings where time-reversal symmetry is not broken (see SI). For both, we estimate the uncertainty of the calibration to be $\pm0.1$~mT.

\section{$\bpar=0$ magnetoresistance peak}

Figure~\ref{fig1}d-j introduces the general phenomenology of the peak in \rb that is the subject of this work.  We focus on one $\{\nu,D\}$ quadrant of the phase diagram, with a correlated insulator at $\nu=2$ and weaker insulating states at $\nu=1$ and 3. (The analogous region at positive $D$ showed a similar behavior but was not studied in detail.)  The metallic state surrounding the $\nu=2$ insulator is sometimes referred to as the `halo' region.  It is distinguished by a higher longitudinal resistance, \rxx, leading to a lighter colour in Fig.~\ref{fig1}d, and defines a region of broken spin and valley degeneracy, leading to values of the reduced Hall filling, $|\nu_H-\nu|$, different from 0 or 4 (Fig.~\ref{fig1}e).  Throughout most of the halo region, the reduced Hall filling is 2, indicating the breaking of one, but not both, of the spin-valley degeneracies; this state is often referred to as a half-metal state, as only half of the original degeneracies remain.  Previous measurements in this device, and in similar devices from other groups, suggest that spin but not valley degeneracy is broken in this half-metal region.\cite{liu2020tunable,cao2020tunable,he2021symmetry, liu2022isospin,liu2023observation}

Above $\nu=3.5$ in the halo, and along its upper-right edge, a reduced Hall filling of 3 is observed, indicating that both spin and valley degeneracies are broken and that the state is a quarter-metal.  Consistent with this observation, a strong AHE appears above $\nu=3.5$ in the halo (circle marker in Figs.~\ref{fig1}e,g). A possible explanation for the much weaker AHE surrounding the $\nu=3$ insulator along the upper edge of the halo (square marker) may be a smaller Berry curvature everywhere except near the top of the band in energy, corresponding to the top of the halo region in gate voltage~\cite{kuiri2022spontaneous}. 

The yellow traces that overlay the map in Fig.~\ref{fig1}d show \rxx recorded while scanning $\bpar$ between $\pm25$~mT at the locations marked by white X's, with the trace at $\nu=3.75,D=-0.47$~V/nm shown in greater detail in Fig.~\ref{fig1}f.  Across the halo and outside of it, most traces are featureless, but strong $\bpar=0$ peaks appear at in-halo locations above $\nu=3.5$, with much weaker peaks along the upper left and lower right boundaries of the halo. A comparison of Figs.~\ref{fig1}d and e supports the observation that the appearance of the $\bpar$ peak is, in almost all cases, correlated with the quarter-metal state with full spin and valley polarization.  The lower left boundary of the halo, where a very small $\bpar$ peak is also observed, does not show evidence of valley polarization in the reduced Hall filling (Fig.~\ref{fig1}e). However, the nearby weak insulating state at $\nu=1$ indicates the tendency toward spontaneous symmetry breaking in that region of the halo, consistent with recent reports of first-order phase transitions and ferromagnetism at the equivalent $\nu,D$ location in an AB-BA analog of t2+2 \cite{liu2023observation}.  In the discussion that follows, we focus on the $\bpar$ peak in gate voltage locations that are clearly valley polarized with robust AHE, that is, in the quarter-metal corner pocket of the halo where $\nu>3.5$.
 
Figures~\ref{fig1}h-j highlight the unusual dependence of this phenomenon on magnetic field orientation.  The extremely anisotropic dependence with respect to in- and out-of-plane fields is most clearly seen in the line traces of  Fig.~\ref{fig1}h, comparing small-field sweeps of $B_x$ (in-plane) and $B_z$ (out-of-plane) taken at the right-most cross in Fig.~\ref{fig1}d.  Similarly, the milliTesla-wide peak at $B_x=0$ is entirely unaffected by $B_z$ up to 10 mT (Fig.~\ref{fig1}j). In contrast, Fig.~\ref{fig1}i shows that the peak is isotropic within the plane, to within experimental resolution.  These characteristics --- in-plane isotropy, extreme in-plane vs.\ out-of-plane anisotropy, and correlation with AHE --- place strong constraints on possible microscopic mechanisms, as discussed in Sec.~\ref{sec:discussion}.

A characteristic of the \rb peak not evident from Fig.~\ref{fig1} is that a given $\bpar$ magnetoresistance measurement is not, in general, repeatable over long periods of time.  Smooth 2D magnetoresistance data like those in Figs.~\ref{fig1}g and h demonstrate that the sample behaviour is sometimes stable enough to remain in a given state for the duration of an hour-long scan, but at other times the sample state jumps abruptly during a scan.  We emphasize that this instability was not a general characteristic of the sample electrostatics: like most van der Waals stacks made up of $hBN$ and graphene, this sample was perfectly stable within experimental resolution away from broken-symmetry states.  Instead, we attribute the irreproducibility of these broken-symmetry features to metastability in the domain structure, and the \rb peak was observed to be the least stable of all broken-symmetry phenomena observed in this sample.

\begin{figure}
	\begin{center}
		\includegraphics [width=1\linewidth]{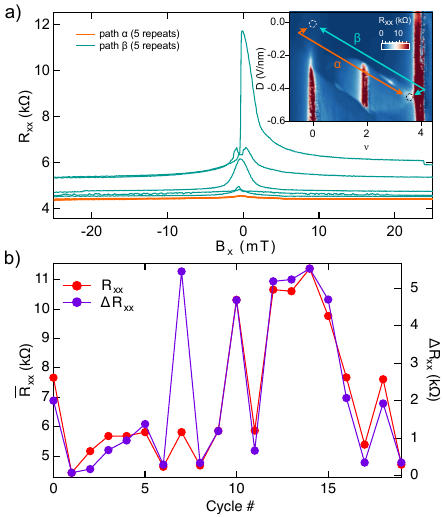}
	\end{center}
	\caption{a) $\rxx(\bpar)$ for five consecutive approaches to the AHE region via each of two gate-voltage trajectories $\alpha$ and $\beta$ (inset), connecting the fully-degenerate metal (white circle, $\{\nu,D\}=\{0,0\}$) to the quarter-metal region (black circle, $\{3.46,-0.44\}$).  Path $\alpha$ never yielded an $\rxx(\bpar)$ peak, whereas path $\beta$ almost always did, with varying peak height.  Approaches were carried out with zero current in the in-plane magnet coils, corresponding to a residual in-plane field of order 0.5~mT.  b) Correlation between the average resistance $\overline{\rxx}$ (red markers) measured over 10~s after arrival via the $\beta$ trajectory, and the peak height $\Delta\rxx\equiv\rxx(\bpar=0)-\rxx(\bpar=10~\mathrm{mT})$ (purple markers) determined by a subsequent $\bpar$ sweep.  Twenty repeats of the $\beta$ approach are shown.}
	\label{fig2}
\end{figure}

\section{Resistive state preparation}

Strong evidence for the domain picture comes from the role that the prior history of sample parameters plays in setting the resistive state.  The $\bpar$ magnetoresistance characteristics were most likely to change if gate voltages were swept out of, then back into, the  state, suggesting that the domain configuration is reset each time the system re-enters the fully symmetry-broken phase.  To demonstrate this, Fig.~\ref{fig2}a compares $\rxx(\bpar)$ behaviour for five consecutive approaches to a point in the AHE region, following two different gate-voltage trajectories labelled $\alpha$ and $\beta$.  Path $\alpha$ reached the AHE region after passing through the half-metal part of the halo, whereas path $\beta$ avoided the halo region, entering the AHE region from the $\nu=4$ insulator.

The protocol for each cycle was as follows: with the magnet at zero, the gates were swept to a fully degenerate position ($\{\nu,D\}=\{0,0\}$), then to the point $\{\nu,D\}=\{3.46,-0.44\}$ in the AHE region following a prescribed trajectory.  $\rxx$ was recorded immediately after arrival, then during a $\pm 25$~mT sweep of $B_x$.  The \bx magnet current was then returned to zero, the gates were swept back to $\{0,0\}$, and the process was repeated 20 times.   The $\{0,0\}$ location was chosen as a reset location in order to eliminate any chance of domain memory within the sample, however, in general it was observed that any departure from the quarter-metal region at the tip of the halo was enough to reset the state.

The $\alpha$ approach path never yielded an $\rxx(\bpar)$ peak state over all 20 repetitions (the first five shown in Fig.~\ref{fig2}a).  The $\beta$ approach frequently did, although the height of the peak varied greatly from cycle to cycle.  The metastable state of the sample could be identified immediately after the gate-voltage approach, without sweeping $\bpar$: the resistance at the point of arrival was consistently higher when the sample was in a state that would exhibit the $\bpar$ peak.  This is because the peak and non-peak states have similar resistance at high $\bpar$, but differ substantially near $\bpar=0$, by as much as several k$\Omega$.  The close correlation between arrival resistance and peak height is shown in Fig.~\ref{fig2}b.  The one datapoint (cycle \#7) for which there is a significant mismatch was one for which the state of the sample jumped discretely during the $\bx$ scan.    

Investigations of several other trajectories taken under similar conditions (data not shown) confirmed a general pattern: approaches through the half-metal consistently yielded non-peak states, while approaches from the $\nu=4$ insulator frequently produced peak states.  It is important to point out, however, that there was a residual in-plane magnetic field at the sample due to trapped flux during these approaches (for example, in  Fig.~\ref{fig2}a,b) even though the in-plane coils of the magnet were set to zero current.  We estimate the true value of $\bx$ during the approach trajectories to have been around $\bx=0.45\pm 0.05$~mT, based the peak locations in the subsequent $\bx$ scans (cf.\ Fig.~\ref{fig2}a); the true value of $\by$ was not measured but was likely similar. In contrast, the true zero of $\bperp$ was calibrated prior to these measurements (see SI) and found to be at a magnet setting of $-1.8$~mT; all Fig.~\ref{fig2} data were collected with the magnet set at this ``true $\bperp=0$'' value.  Even though the residual approach fields were small, in the discussion that follows we will see that they played an critical role in the appearance of the data.  

\begin{figure}
	\begin{center}
		\includegraphics [width=1\linewidth]{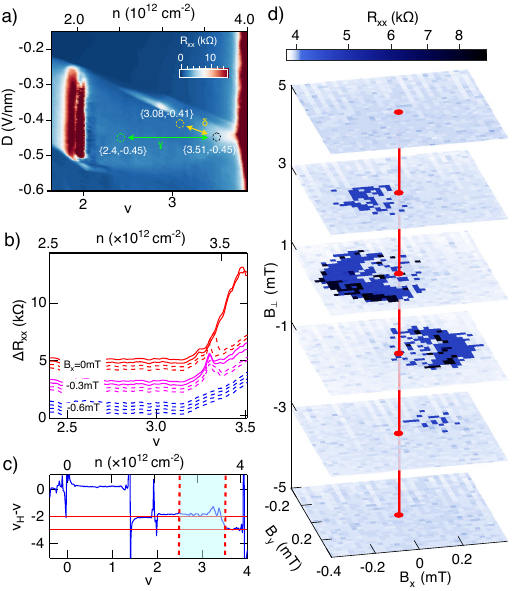}
	\end{center}
	\caption{a) $\rxx$ map of the tip of the halo region where the $\bpar=0$ MR peak is observed, with gate-driven trajectories $\gamma$ (green) and $\delta$ (yellow) used for panels (b) and (d); starting and ending points are labelled in terms of $\{\nu,D\}$.  b)  $R_{xx}(\nu)$ traces along $\gamma$ at three values of $B_x$, with $\bperp$ and $B_y$ set to zero. Solid (dashed) lines indicate traces terminating in the large-peak (small-peak) state.  (c) Reduced Hall filling, $|\nu_H-\nu|$, measured along the full range $\nu=0$ to 4 at constant $D$; the light blue shading highlights the portion corresponding to trajectory $\gamma$.  (d) Terminal $\rxx$ after ramping along $\delta$ in (a) for various $\vec{B}$.}
	\label{fig3}
\end{figure}

Figure~\ref{fig3}a shows two more gate voltage trajectories, $\gamma$ and $\delta$, that arrive at the quarter-metal state at $\nu=3.51,D=-0.45$~V/nm from within the half-metal state.  We consider first trajectory $\gamma$, tracking a line of constant $D=-0.45$~V/nm from $\nu=2.4\rightarrow 3.51$.  Figure~\ref{fig3}b shows $\rxx$ traces measured along that trajectory at three different in-plane magnetic fields. Independent of field, the resistance starts low (around 1~$k\Omega$) at $\nu=2.5$, a location in gate-space that we believe to be a spin-polarized half-metal as described in Section~III, consistent with a reduced Hall filling $|\nu_H-\nu|=2$ at that point (Fig.~\ref{fig3}c).  The resistance remains nearly constant until $\nu=3.2$, at which point it begins to rise.  Once $\nu$ passes 3.4, the resistance trace bifurcates into two trajectories, with the probability of each depending on the precise value of the in-plane magnetic field during the gate sweep.  This bifurcation is concomitant with a step in $|\nu_H-\nu|$ to $3$ (Fig.~\ref{fig3}c), signaling the onset of the spin- and valley-polarized quarter metal state.  At $\bx=0$, there is approximately a 50\% chance of the sample reaching the high-R state, while at $\bx=0.6$~mT the sample consistently enters the low-R state.

Figure~\ref{fig3}d maps the resistance of the sample in the quarter-metal state as a function of the magnetic field applied during the gate-voltage ramp used to reach it.  Here, we follow trajectory $\delta$ in Fig.~\ref{fig3}a for mT-scale magnetic field vectors along all spatial directions, and display the resistance upon arrival at $\nu=3.51,D=-0.45$~V/nm.  We note that the pattern in Fig.~\ref{fig3}d, where the sample is reset by a gate-voltage sweep at each value of $\vec{B}$, is qualitatively different from the \rb peak discussed in Section~III, which is measured by sweeping $\bpar$ at fixed gate voltage after the sample has already entered the ordered phase.  Unlike the \rb peak, the pattern in Fig.~\ref{fig3}d is not isotropic in $\bpar$: when $B_\perp$ is positive, the large-peak state appears mostly for positive $\bx$, and vice versa.  The maximum resistance of the large-peak state is not found at exactly $\bpar=0$, but rather at $\pm~0.3$~mT depending on the sign of $\bperp$.  Finally, the large-peak state disappears for gate-voltage sweeps carried out when $|\bpar|\gtrsim 0.4$~mT, but persists up to values of $|\bperp|$ a factor of 6 larger.  The range of $\bpar$ explored here is significantly narrower than the 1--2~mT width of the \rb peak (see, e.g., Fig.~\ref{fig1}g).  Low resistance values away from $B=0$ in this panel therefore indicate that the sample entered the small-peak state, not that it was in the large-peak state but measured off-peak.

\begin{figure}
	\begin{center}
    \includegraphics [width=1\linewidth]{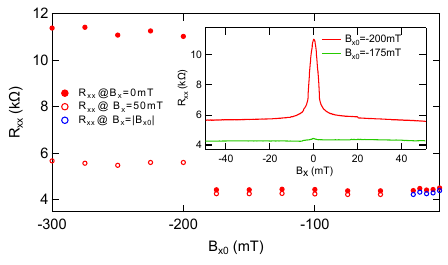}
	\end{center}
	\caption{Re-emergence of the large-peak state at large approach fields.  Peak (solid) and off-peak (open) values of $\rxx$, after following approach trajectory $\alpha$ from Fig.~\ref{fig2}a with an in-plane approach field $B_{x0}$ ranging from $-5$ to $-300$~mT.  At the final gate-voltage location, $B_x$ is swept between $\pm 50$~mT, or $\pm B_{x0}$ for $B_{x0}<50$~mT, and the peak and off-peak resistances are extracted.  For $|B_{x0}|\leq 175$~mT, the two values are nearly equal (no $\bpar$ peak), but for $|B_{x0}|\geq 200$~mT a large peak re-emerges.  Inset: representative $\rxx(B_x)$ sweeps at $B_{x0}=-200$~mT (large peak) and $B_{x0}=-175$~mT (no peak).}
	\label{fig4}
\end{figure}

The data in Fig.~\ref{fig3}d, by themselves, would imply that the sample always locks into the small-peak state when the gate-voltage approach to the quarter metal takes place with $\bpar$ above a fraction of a mT.  We find, however, that the large-peak state re-emerges when the in-plane approach field is larger than $\sim 175$~mT.  Figure~\ref{fig4} shows data taken along trajectory $\alpha$ from Fig.~\ref{fig2}a---corresponding to the half-metal to quarter-metal approach---but now with approach fields ranging from $\bx=-5$ to $-300$~mT.  At the final gate-voltage location (black dot in Fig.~\ref{fig2}a), $\bx$ is swept through zero while $\rxx$ is recorded, and the peak and off-peak resistances are extracted (solid and open circles, respectively).  For approach fields at or above 5~mT but below $\sim 175$~mT, the sample resistance is low (near 4~k$\Omega$) both on and off peak (green curve in inset).  Above 175~mT, the on-peak resistance increases by nearly a factor of 3, to 11.5~k$\Omega$, while the off-peak resistance rises to 5.5~k$\Omega$ (red curve in inset).

\section{Discussion}
\label{sec:discussion}

The data presented here establish several facts about the anomalous $\bpar=0$ resistance peak in t2+2.  First, the peak is associated with magnetic domains: its presence and strength depend on the history of the sample, including the gate-voltage trajectory used to enter the ordered state and the magnetic field present during that trajectory (Figs.~\ref{fig2}--\ref{fig4}).  Second, these domains are domains of valley polarization: the $\bpar$ peak appears exclusively at gate voltages where the anomalous Hall effect signals spontaneous time-reversal symmetry breaking and previous transport data is consistent with domains~\cite{kuiri2022spontaneous,liu2023observation}. We note that domain structures of orbital magnetization have been directly visualized by scanning magnetometry in magic-angle twisted bilayer graphene~\cite{grover2022chern, tschirhartImagingOrbitalFerromagnetism2021a}, though not yet to our knowledge in twisted double bilayer graphene.  Third, spin is involved in addition to valley.  The $\bpar$ peak is isotropic within the graphene plane (Fig.~\ref{fig1}i), whereas an orbital or lattice-related effect would be expected to reflect graphene's $C_3$ symmetry.  The peak is sensitive to $\bpar$ but essentially unaffected by $\bperp$ up to at least 10~mT (Fig.~\ref{fig1}j), indicating that the relevant energy scale is set by spin Zeeman coupling rather than by orbital coupling to $\bperp$.  The Zeeman energies implied by the milliTesla width of the peak are far below $k_BT$, ruling out conventional quasiparticle scattering.  Similarly, the domain configuration depends more strongly on $\bpar$ than on $\bperp$ during formation (Fig.~\ref{fig3}d).

The simultaneous involvement of spin and valley is naturally explained by the Kane-Mele-type spin-orbit interaction present in bilayer graphene and therefore presumably in t2+2~\cite{kane2005quantum,konschuh2012theory,kurzmann2021kondo,jing2025electric,adam2025entropy}.  This interaction locks valley polarization to out-of-plane spin: $K$ with $\uparrow_z$ and $K'$ with $\downarrow_z$, or vice versa.  Its strength in bilayer graphene is around 50--100~$\mu$eV~\cite{konschuh2012theory,banszerus2020observation,kurzmann2021kondo,jing2025electric,adam2025entropy}, weak compared to many materials but dominant at the temperatures and fields relevant here.  As a consequence, domains of valley polarization are simultaneously domains of spin polarization, with spins in adjacent domains that may be anti-aligned.

These observations raise two distinct questions.  For a given domain configuration, what is the mechanism by which $\bpar$ near zero produces excess resistance?  And how is the domain configuration itself set by the magnetic field present when the ordered state forms?  In both cases $\bpar$ plays a central role, but the physics is different: the first concerns transport across existing domain walls, while the second presumably concerns the energetics of domain nucleation.

\textit{Mechanism of the $\bpar$ peak.}---Although the data in this paper are not sufficient to identify the microscopic origin of the resistance peak unambiguously, we note one speculative scenario that is consistent with all of the symmetry constraints listed above, described in more detail in \cite{ray1_3}.  If adjacent domains carry opposite spin polarizations, a transport current flowing across a domain wall will exert a spin-transfer torque on the magnetic moments within the wall.  For a wall between domains with opposite out-of-plane polarization, the moments within the wall are expected to lie in the graphene plane when Hunds coupling is stronger than spin-orbit coupling, as is likely the case here\cite{das2026multicomponent}. At $\bpar=0$, in-plane rotations of these moments are only weakly pinned by (Rashba type) spin-orbit coupling that break the $U(1)$ spin rotation symmetry about the $z$ direction. Consequently, even a tiny applied current can generate sufficient torque to drive steady-state precession.

In 1986, Berger~\cite{berger1986possible} predicted that precession of a spin texture generates a time-varying Berry-curvature flux that acts as a topological electromotive force~\cite{yang2009universal} --- an effect sometimes called the ferro-Josephson effect by analogy with the DC Josephson effect in superconductors.  This would manifest as excess longitudinal resistance at $\bpar=0$.  A small finite $\bpar$ would break the in-plane rotational symmetry, pinning the domain-wall moments and blocking precession until the current-induced torque exceeded a critical value set by the Zeeman energy.

Were such a mechanism at work, the extreme sensitivity to $\bpar$ would follow from the small Zeeman energy needed to pin in-plane domain-wall moments; the isotropy within the plane would reflect the fact that only the magnitude of $\bpar$, not its direction, sets the pinning strength; and the insensitivity to $\bperp$ would be expected because the domains are already polarized out of plane by exchange and spin-orbit coupling, so that moderate $\bperp$ shifts the relative energy of the two domain orientations but leaves the domain-wall structure approximately intact.  This scenario also predicts that the excess resistance should depend strongly on the measurement current, vanishing in the limit of zero bias where there is no torque to drive precession.  The instability of the domain structure in this sample made a systematic study of current dependence impractical, but in the few measurements where it was attempted, the $\bpar$ peak did show a strong current dependence consistent with this expectation (see supplementary material).

\textit{Domain configuration.}---A separate question is why the domain structure depends so sensitively on $\bpar$ during formation when arriving from the half-metal state.  The data in Figs.~\ref{fig3}d and \ref{fig4} show that the resistive state of the quarter-metal depends on the magnetic field that was present when gate voltages were swept into the ordered phase.  The domain pattern is evidently not set by equilibrium at the final gate-voltage location but is frozen in during the transition.  Even a fraction of a milliTesla during this transition is enough to select between domain configurations that produce qualitatively different transport.  As with the resistance peak itself, $\bpar$ is more important than $\bperp$ in setting the domain configuration (Fig.~\ref{fig3}d), pointing again to spin rather than valley as the degree of freedom most directly affected by the field during formation.

Although the microscopic origin of the $\bpar$-dependent domain configurations is not yet understood, a possible clue lies in the magnetic properties of the parent state.  The gate-voltage trajectories in Figs.~\ref{fig3} and \ref{fig4} approach the quarter-metal from the half-metal, which is valley-unpolarized but spin-polarized with the polarization direction mostly in-plane when Hund's coupling dominates Kane-Mele spin-orbit coupling, as expected for our sample.  In the half-metal state, even small magnitudes of $B_{\parallel}$ could then select a preferred in-plane orientation, suppressing slow variations of the spin direction across the sample. Without a concrete mechanism in mind, we simply point out the possibility that the resulting spin configuration in the half-metal could plausibly seed the valley domain structure when the quarter-metal forms.

Taken together, the data presented here establish that the anomalous $\bpar=0$ resistance peak in the t2+2 quarter-metal depends on magnetic domain configuration, and that this configuration is frozen in by the magnetic field present during the phase transition from the half-metal.  The role of $\bpar$ is twofold: it influences the domain structure during formation, and it controls the resistance of existing domain walls, possibly through suppression of current-driven precession of domain-wall moments.  In both cases, the involvement of spin --- coupled to the valley degree of freedom by spin-orbit interaction --- appears essential.  A complete microscopic theory connecting these observations remains an open challenge.

\section{Acknowledgements}
The authors thank Allan MacDonald, Mona Berciu, Marcel Franz, and Stuart Parkin for helpful discussions, and Silvia Folk for helpful experimental contributions. Experiments at UBC were undertaken with support from the Natural Sciences and Engineering Research Council of Canada; the Canada Foundation for Innovation; the Canadian Institute for Advanced Research; the Max Planck-UBC-UTokyo Centre for Quantum Materials and the Canada First Research Excellence Fund, Quantum Materials and Future Technologies Program; and the European Research Council (ERC) under the European Union's Horizon 2020 research and innovation program, Grant Agreement No. 951541. C.H is supported by NSF CAREER grant award No.~DMR-2541471 and the Department of Energy, Office of Basic Energy Sciences under Award No.~DE-SC0024346. K.W. and T.T. acknowledge support from the JSPS KAKENHI (Grant Numbers 21H05233 and 23H02052) and World Premier International Research Center Initiative (WPI), MEXT, Japan.

\bibliography{main}

\clearpage
\widetext
\begin{center}
\textbf{\large Supplemental Materials: Metastable magnetic domains and the anomalous $B_\parallel=0$ resistance peak in twisted double bilayer graphene}
\end{center}
\setcounter{equation}{0}
\setcounter{figure}{0}
\setcounter{table}{0}
\setcounter{section}{0}
\setcounter{page}{1}
\makeatletter
\renewcommand{\theequation}{S\arabic{equation}}
\renewcommand{\thefigure}{S\arabic{figure}}
\renewcommand{\thesection}{S\arabic{section}}
\renewcommand{\bibnumfmt}[1]{[S#1]}
\renewcommand{\citenumfont}[1]{S#1}
\makeatother

\section{$\bperp$ calibration}
\begin{figure}[ht]
    \centering
    \includegraphics[width=0.5\linewidth]{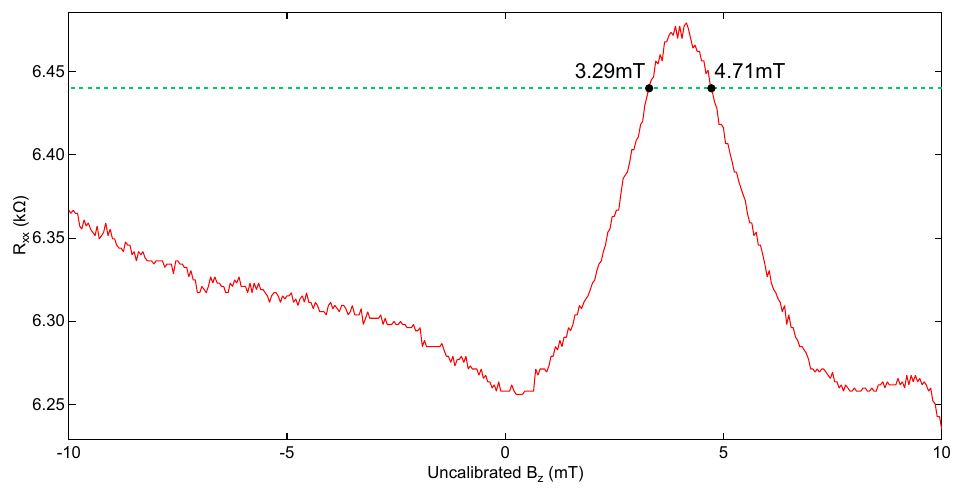}
    \caption{Calibration of $\bperp=0$ using weak localization.  At gate voltages where the four-fold band degeneracy is intact, $\rxx(\bperp)$ shows a weak localization peak centered at $\bperp=0$.  The peak center determines the true field zero, accounting for trapped flux in the magnet coils.  In this example, the peak is centered at a coil setting of $4.0$~mT.}
    \label{fig:calibrate_perp}
\end{figure}

\section{Highly non-linear character of $\bpar=0$ peak}

\begin{figure}[ht]
	\begin{center}
		\includegraphics [width=1\linewidth]{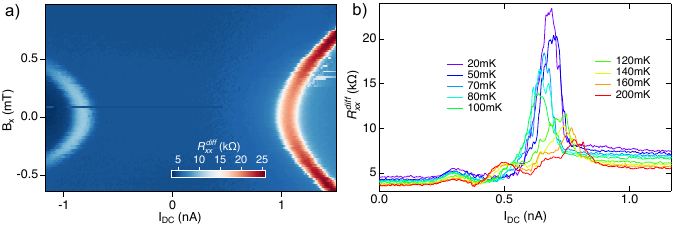}
	\end{center}
	\caption{Current and temperature dependence of the $\bpar=0$ resistance peak.  (a) Differential resistance $dV_{xx}/dI$ as a function of DC bias current $I_{DC}$ and in-plane field $\bx$, measured by adding $I_{DC}$ to a 100~pA (rms) AC excitation and locking in to $V_{xx}$.  For each value of $\bx$, $I_{DC}$ was scanned before stepping $\bx$ to its next value.  The $\by$ and $\bperp$ coils were set to zero current; the true field zero was not calibrated for these measurements, but we estimate residual $\by\sim 0$ and $\bperp\sim 5$~mT from the magnet history.  (b) Temperature dependence of $dV_{xx}/dI$ vs $I_{DC}$, with all fields calibrated to zero, collected several days later during a brief period of sample stability.  The AC excitation was reduced to 30~pA to avoid obscuring fine features.}
	\label{figs2}
\end{figure}

The speculative mechanism discussed in the main text --- current-driven precession of domain-wall moments --- predicts that the $\bpar$ peak should depend strongly on bias current: precession requires a current-induced torque exceeding the pinning set by Zeeman energy, so the excess resistance should appear only above a threshold current that grows with $|\bpar|$.  We cooled the sample a second time to investigate this prediction.  The domain state was extremely unstable during this cooldown, and Fig.~\ref{figs2}(a) shows the only 2D scan completed without the state jumping mid-measurement.

At low DC bias ($I_{DC}$ below $\sim 1$~nA), the differential resistance is flat and featureless.  Near $\bx=0$, a sharp peak in $dV_{xx}/dI$ appears at $I_{DC}\approx 1$~nA, then drops back to the flat background at higher bias.  As $|\bx|$ increases, the current at which this peak occurs grows, reaching $\sim 1.5$~nA at the edges of the scan ($\bx=\pm 0.6$~mT).  This pattern is qualitatively consistent with a threshold current that increases with $|\bpar|$.  This observation offers a natural explanation for the $\rxx(\bpar)$ peak in the main text.  The AC bias current used there, $I_{ac}=1.4$~nA (rms), has a peak amplitude of $1.4\sqrt{2}\approx 2$~nA.  At $\bpar=0$, the sharp peak in $dV_{xx}/dI$ lies within the range of currents sampled during each AC cycle, so the lock-in registers an elevated $\rxx$.  As $|\bpar|$ increases, the $dV_{xx}/dI$ peak shifts to higher $I_{DC}$; once it moves beyond the AC amplitude, the lock-in no longer samples it and $\rxx$ drops back to its background value.  The $\bpar$ peak in the main-text measurements is thus a direct consequence of the current-dependent features revealed here.

Panel (b) shows the temperature dependence at $\bpar=0$: the peak in $dV_{xx}/dI$ appears at a lower current ($\sim 0.7$~nA, likely because all fields were calibrated to zero for this measurement) and is sharpest and tallest at the base temperature of 20~mK.  It broadens and diminishes with increasing temperature, becoming very weak by 200~mK.  

\end{document}